\documentclass[conference]{IEEEtran}
\IEEEoverridecommandlockouts
\usepackage{spconf,amsmath,graphicx}
\usepackage{balance}
\usepackage{paralist}

\title{Interpreting deep urban sound classification in driving environment using Layer-wise Relevance Propagation}
\name{Marco Colussi and Stavros Ntalampiras\thanks{We  would  like  to  thank  NVIDIA  Corporation for the donation of a Titan V GPU used in this research.}}
\address{University of Milan \\ Department of Computer Science \\ via Celoria 18, Milan, Italy}
\begin{document}
%
\maketitle
\begin{abstract}
Explaining the decision-making process of modern AI-based systems is a challenging task especially when deep neural architectures are involved. After constructing a deep neural network (DNN) for urban sound classification, this work focuses on the sensitive application of assisting drivers suffering from hearing loss. As such, clear etiology justifying and interpreting model predictions comprises a strong requirement. To this end, we used two different representations of audio signals, i.e. Mel and constant-Q spectrograms, while the decisions made by the DNN are analyzed via layer-wise relevance propagation. At the same time, frequency content assigned with high relevance in both feature sets, indicates extremely discriminative information characterizing the present classification task. Overall, we present an etiology framework for understanding deep urban sound classification.
\end{abstract}
\section{Introduction}
\label{sec:intro}
Over 5\% of the world’s population, nearly 466 million people, live with hearing loss \cite{hearing-loss} and are identified as deaf or hard of hearing (DHH). In their everyday lives, they might experience severe difficulties especially when carrying out complex tasks, such as driving a vehicle.

Despite visual modality being the most important one for the driving experience, the auditory sense is still remarkably relevant. Several studies have been conducted on increasing sound-awareness including a sound-detecting \cite{10.1145/2982142} mobile application that helps DHH individuals in having available tactile and visual feedback notifications for different stimuli. 
The ever-increasing relevance of audio-awareness is demonstrated by the extensive amount of research carried out in this field \cite{Ntalampiras2014,9103031,9053274}.

Moving to a driving environment and seeing the problem from a DHH individual point of view, we can easily imagine that auditory evidence  providing information about approaching vehicles, hazards, potential vehicle problems, or appearance of emergency vehicles, can be considered vital since it is  almost impossible to perceive them based on other senses. The specific problem belongs to the computational auditory scene analysis \cite{Brown1994} field where the goal is to use audio coming from the surrounding environment in order to achieve scene understanding, i.e. recognize potentially overlapping in time sound events and/or soundscapes. The analysis of the related literature suggests that using deep neural networks (DNN) operating on spectrogram-like representations of audio signals is one of the most prominent and suitable ways for classifying audio signals \cite{7952265,li2020neural,8678825,9132743}.

Unfortunately, even though DNNs are able to provide excellent performance, both their decisions and analysis are not directly interpretable, often making their application in sensitive real-world tasks problematic. In fact, due to their complex structures, DNNs are often considered as black-boxes which do not provide additional information regarding the cause behind a specific classification decision, such as which features are highlighted by the neural network \cite{yosinski2015understanding}. 
The closest paper to this work is presented in \cite{becker2018interpreting} where the authors attempt to explain DNN's predictions in a speech classification task. Considering the present application, availability of a rigorous etiology framework focusing on explainable AI is of paramount importance. This will allow us not only to rely on the accuracy of the choices made by the intelligent system, but also to meaningfully relate input with output data.

Both understanding and interpretation of decisions are therefore essential to create a reliable system \cite{8979157}; indeed, such a feature will be considered a standard requirement in machine learning based solutions according to the European Commission \cite{EUAIreport}. However, to this date, DNN operation is not fully explainable, thus this work focuses on investigating, analyzing and studying the network and its decisions.

More in detail, this article details a robust model classifying environmental sounds typically surrounding a vehicle. Subsequently, we provide an effective analysis of the results provided by our model based on the Layer-wise Relevance Propagation heatmaps method \cite{10.1371/journal.pone.0130140}. More specifically, we operate on two feature sets characterizing the available sounds, i.e. Mel and constant-Q spectrograms which provide diverse information and representation of the audio signals. Subsequently, we provide clear insights regarding why and how the proposed model makes predictions. At the same time, we reveal highly distinctive frequency content which was identified as relevant in both representations. Towards a reproducible research, our experiments were carried out using the Urban8K dataset \cite{Salamon:UrbanSound:ACMMM:14} which is publicly available along with our code which can be found at \url{https://github.com/warpcut/LRP-project}. The final set of considered classes used to train the network and analyse its prediction is composed of the following: \textit{air conditioner}, \textit{car horn}, \textit{children playing}, \textit{dog bark}, \textit{drilling}, \textit{engine idling}, \textit{gun shot}, \textit{jackhammer}, \textit{siren}, \textit{street music}.

The rest of this work is organized as follows: section \ref{sec:problem_form} formulates the problem. Section \ref{sec:model} analyses the classification model and section \ref{sec:features} describes the employed feature sets. Section \ref{sec:lrp} outlines the layer-wise relevance propagation techniques, while section \ref{sec:experiments} includes the experimental set-up and results. Finally, section \ref{sec:conclusions} draws our conclusions.


\begin{figure}[t]
	\centering
	\includegraphics[width=0.52\textwidth,trim=100 0 0 0]{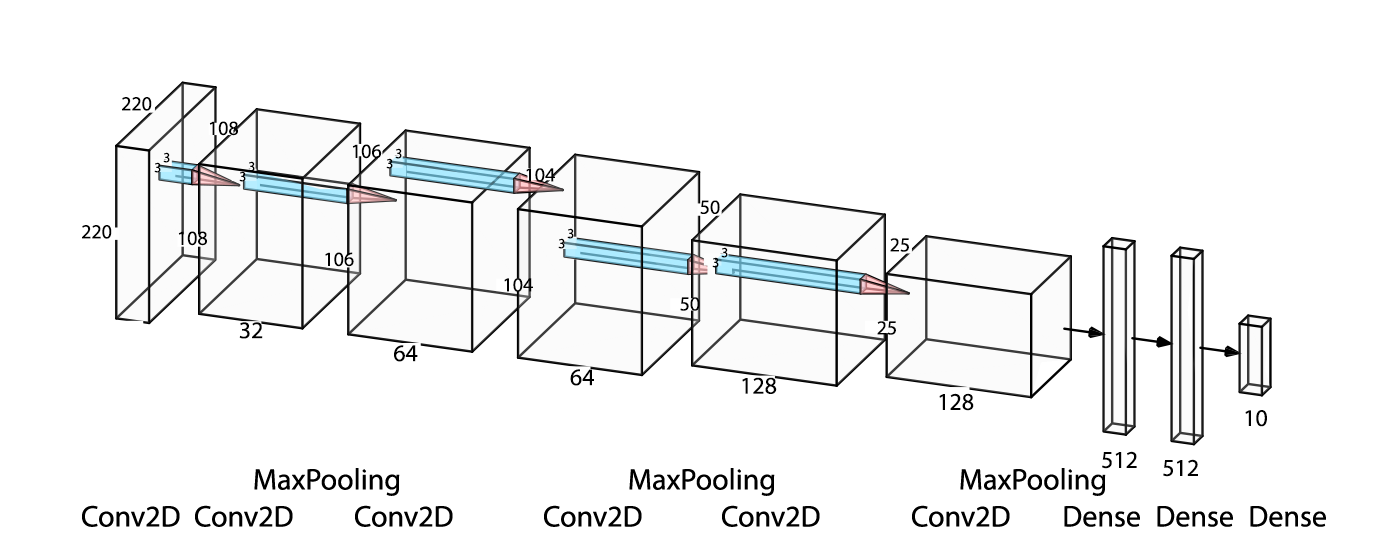}
	\caption{The architecture of the considered audio classification model.}
	\label{fig:alex-net}
\end{figure}

\section{Problem formulation}
\label{sec:problem_form}
The present problem definition assumes availability of a monophonic audio signal $y$ the content of which characterizes urban environmental sound events in set of classes $\mathcal{D}$. 
$\mathcal{D}=\{C_1,\dots,C_n\}$ and stationary over time, i.e. composed of known classes both during training and testing phases. Then, as typically assumed in generalized sound recognition systems \cite{Ntalampiras2019jaes}, each sound event follows a consistent and unknown probability density function $P_i, i\in[1,m]$,

Lastly, we suppose that a training set is available, denoted as $TS=y_t,t\in[1,T_0]$  with annotated pairs $(y_t,C_i)$, where $t$ is the time instant and $i\in[1,m]$.
The final aim is to identify urban environmental sound events in novel audio sequences while assessing the relevance assigned by the classifier along the frequency content of each class.

\section{Classification model}
\label{sec:model}
In this section, we describe the model selected following the state-of-the-art for audio classification tasks, i.e. a Convolutional neural network (CNN).
CNN comprises one of the most suitable machine learning models for carrying out multi-class image classification tasks \cite{8521242, 8681654, Ntalampiras2021}. Their main characteristic is their specialization capacity in capturing local features in images highlighting existing spatial relationships.

The CNN considered in this work is based on AlexNet \cite{alexnet} architecture originating from the image recognition domain. It is composed of five convolutional layers followed by downsampling performed by max-pooling layers; the network is completed by three fully connected layers for the classification (see Fig. \ref{fig:alex-net}. Slight alterations were applied to the original implementation of AlexNet mainly to adapt it to the present dataset. The first convolutional layer takes as input an image of size 220x220x3, with a channel last input data with the filter size of 32 and stride of 3x3. It should be noted that while the stride remain equal for all following layers, the filter dimensionality is doubled every 2 layers.

The three fully connected layers are made by two 512 dense layers and the final one is sized so as to generate ten-class predictions as we can see in Fig\ref{fig:alex-net}. To prevent overfitting and stabilize the model, we added a dropout layer with a 0.4 probability rate and we introduced batch normalization layers after the convolutional ones. Finally, the dataset has been randomly divided in three separate sets, one for training ($70\%$), one for validation ($20\%$) and one for test ($10\%$).

\begin{figure}[t]
	\centering
	\includegraphics[width=0.5\textwidth,trim=70 0 0 0]{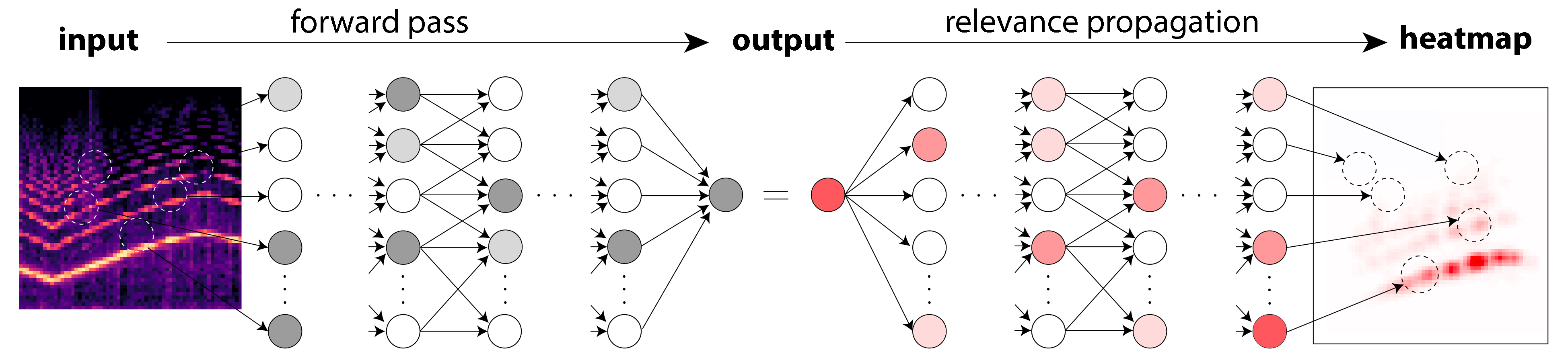}
	\caption{Generation of the layer-wise relevance propagation map.}
	\label{fig:hm-generation}
\end{figure}



\section{Feature sets}
\label{sec:features}
The learning algorithm described in section \ref{sec:model} was trained and tested on $220x220x3$ images representing spectrograms of single channel audio samples. We employed two diverse spectrogram types in order to investigate their respective relevant parts and identify whether specific frequency content is equally relevant/irrelevant. To this end, we employed Mel and constant-Q spectrograms, which are briefly explained next.

\begin{figure*}[t]
	\includegraphics[width=\textwidth,trim=0 0 00 0]{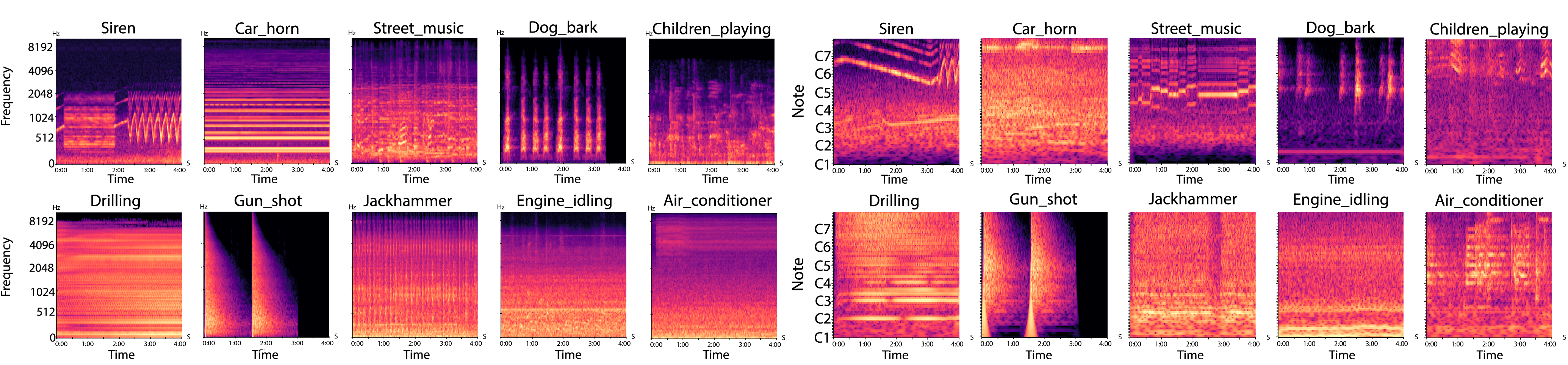}

	\caption{Spectrograms representing every class available in the Urban8K dataset \cite{Salamon:UrbanSound:ACMMM:14}: \textit{air conditioner}, \textit{car horn}, \textit{children playing}, \textit{dog bark}, \textit{drilling}, \textit{engine idling}, \textit{gun shot}, \textit{jackhammer}, \textit{siren}, \textit{street music}.}
	
	\label{fig:all-spect}
\end{figure*}

\subsection{Mel-spectrogram}
Mel-spectrograms, demonstrated in Fig. \ref{fig:all-spect}, comprises one of the most popular representation for audio signals \cite{Salamon2015eusipco}. They consist in a frequency domain representation, where on the x-axis we have the time, and on the y-axis, we decompose the magnitude of the signal into its components, i.e. the frequencies in the Mel-scale. More information is available in \cite{8678825,9132743}.

\begin{figure}[b]
	\includegraphics[width=0.5\textwidth,trim=10 0 0 0]{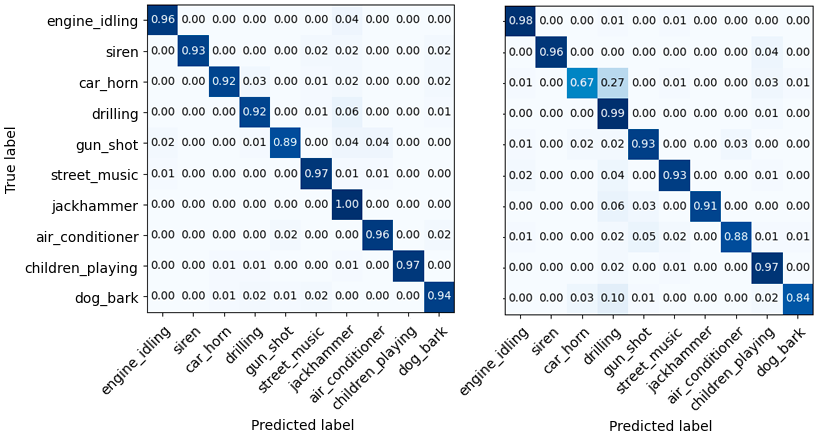}
	\caption{Confusion matrices of the model when trained with Mel (left part) and constant-Q spectrograms (right part).}
	\label{fig:confusion-matrices}
\end{figure}

\subsection{Constant-Q spectrogram}
In addition, we investigated the spectrogram based on the constant-Q transform shown in Fig.\ref{fig:all-spect}. It includes geometrically-spaced center frequencies, while time resolution increases towards higher frequencies, a process mimicking the human auditory system. Such a generation process results in different patterns w.r.t Mel-spectrograms and permits a comparable analysis of discriminative spectral areas as they are identified via the layer-wise relevance propagation technique.


\section{Layer-wise Relevance Propagation}
\label{sec:lrp}
The layer-wise relevance propagation (LRP) technique \cite{10.1371/journal.pone.0130140} aims at detecting the most significant input features for a given prediction, thus it may provide insights regarding the reasons leading to both successful and erroneous predictions.

LRP decomposes the predictor’s output into relevance scores $R_i$'s, associated with node $i$. This is carried out by iterating from the output layer to the input one, redistributing $R_i$ from the upper layer to the lower one in accordance with the contribution on the activation of the input to the output neuron, preserving the total relevance from a layer to another.

The relevance scores are contained in a heatmap $R(x) = {R_p(x)}$, where $p$ must be conservative and positive \cite{10.1371/journal.pone.0130140}. The rule of propagation presented in \cite{DBLP:journals/corr/MontavonBBSM15} is based on deep Taylor decomposition, i.e. applying the decomposition to the subfunctions that compose the DNN and redistributing the relevance to the original pixel level, thus forming the heatmap.

Let $w_j$ denote the vector of weights associated with  parameters connecting the input to the neuron $x_j$. Relevance is redistributed based on the square magnitude of $w_j\forall i,j$ as follows:
\begin{equation}
\label{eq:wsquare}
R_i = \sum_j{w^2_{ij}\over \sum_{i}w^2_{ij}}R_j
\end{equation}
where $R_j$ is the relevance associated with neuron $x_j$ existing in the upper layer. $R_i$ is computed by adding the scores of the upper levels to which neuron $x_i$ contributes: $R_i=\sum_j R_{i\gets j}$.

\subsection{The LRP flat rule}
Towards a data-independent Eq. \ref{eq:wsquare}, we adopted the LRP flat rule which was introduced in \cite{DBLP:journals/corr/abs-1902-10178}. As such, all weights present in Eq. \ref{eq:wsquare} are set to one, leading to
\begin{equation}
\label{eq:flatrule}
R_i = \sum_j{1\over \sum_i 1}R_j
\end{equation}
This produces a decomposition that is both model- and data-independent by projecting uniformly the relevance to the lower layer. The overall process is depicted in Fig. \ref{fig:hm-generation}.

\begin{figure*}[t]
	\includegraphics[width=\textwidth]{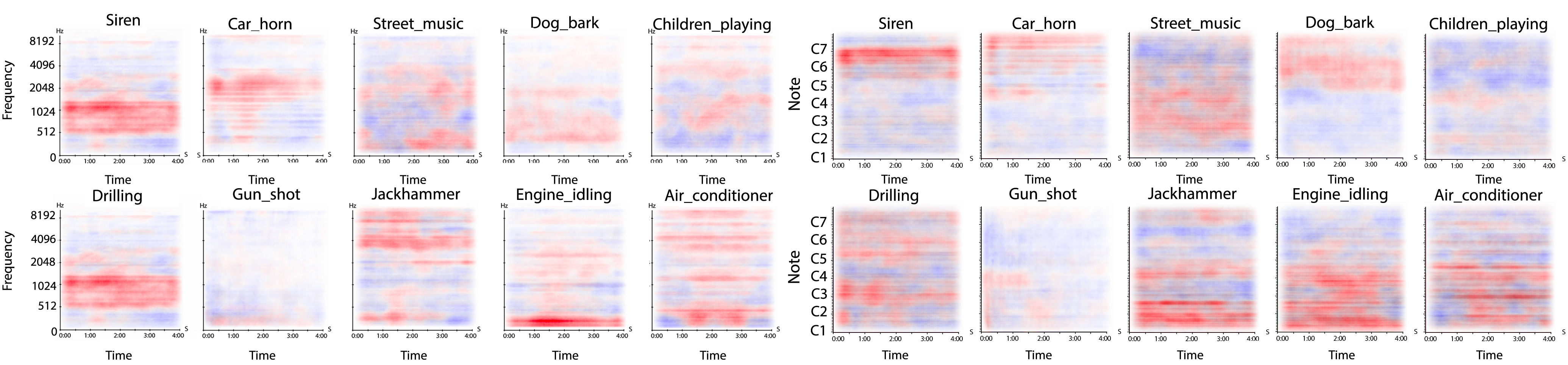}
	\caption{Qualitative evaluation of relevance maps for every sound class considering both Mel (left part) and constant-Q (right part) spectrogram representations.}
	\label{fig:avg-spect}
\end{figure*}

\begin{figure}[b]
	\includegraphics[width=0.48\textwidth,trim=0 20 5 0]{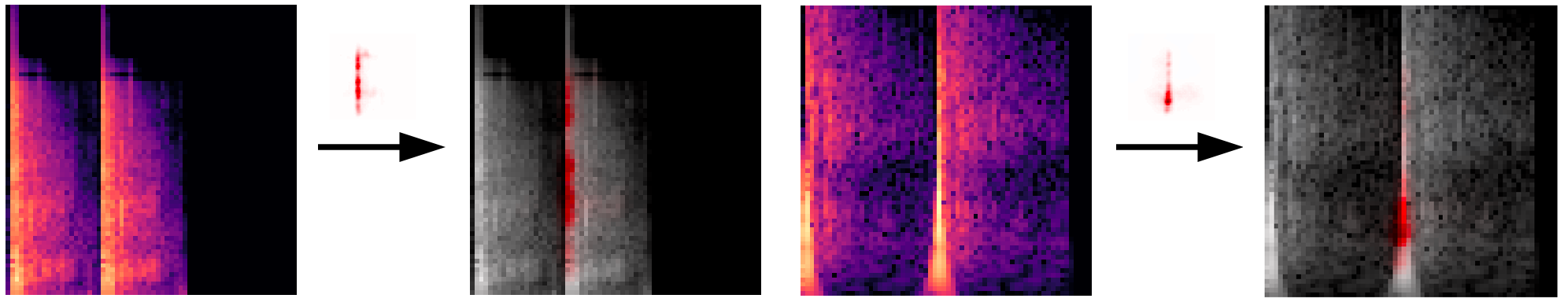}
	\caption{LRP-based analysis of Mel and constant-Q based spectrograms extracted from two different gunshot audio samples. A thumbnail of the relevance map corresponding to each sample is given over the arrow.}
	\label{fig:gun_mel}
\end{figure}

\section{Experimental set-up and results}
\label{sec:experiments}
This section describes the experimental setup and presents the results of the outlined analysis method. More specifically, it presents 
\begin{inparaenum}[a)]
	\item the dataset,
	\item model parameterization in terms of training and evaluation,
	\item classification results for both feature sets, and
	\item prediction analysis results based on the LRP technique.
\end{inparaenum}

\subsection{Dataset}
We employed the UrbanSound8K \cite{Salamon:UrbanSound:ACMMM:14} dataset, which is an open-source and freely available dataset containing 8732 audio excerpts belonging to 10 different classes mentioned in sec. \ref{sec:intro}. The specific dataset was used since it encompasses the most significant classes while considering a driving-like environment. Even though not all considered classes are essential, using every available data could help the model in generalizing when used in unrestricted environments including \textit{a-priori} unknown sound events. At the same time, it boosts reproducibility of the current experiments.


During data preprocessing, we homogenize the audio samples in terms of length and sampling rates so as to extract spectrograms of consistent sizes facilitating the CNN. Thus, when necessary, we resampled them to 44.1KHz and applied trimming/padding so as to reach a 4 seconds length. It should be mentioned that we employed the audio analysis python package Librosa \cite{mcfee2015librosa}.

\subsection{Training and evaluation parameterization}
Training was performed using an early stopping callback on a total of 80 epochs; patience parameter was set to 10, while the process regarding both models stopped at approximately 60 epochs. The best-performing sets of weights were then restored forming the final models.
We used Nesterov Adam optimizer \cite{Zhong2020}, and to prevent overfitting, we designed a weighting scheme for the available classes were assigning higher importance when encountering under-represented samples\footnote{This was achieved by passing a weight dictionary to the \emph{fit()} keras method.}.

We used loss and accuracy as evaluation metrics and categorical crossentropy as loss function. Finally, we obtained two models with more than satisfactory accuracy reaching more than $95\%$ for the Mel spectrogram, and close to $90\%$ for the constant-Q one.

The achieved recognition rates per class are tabulated in the normalized confusion matrices shown in Fig. \ref{fig:confusion-matrices}. We observe that every class is recognized with excellent rates. In both cases, the models perform accurately on unseen data, with quite high precision and recall scores ($\approx95\%$). Therefore, we can proceed with the LRP-based analysis of the model itself and its predictions. Such observations regarding achieved performance are in line with the related works reported in \cite{8635051,Salamon2017,Jindal2020}.

\subsection{Analysis and relevance map extraction}
\label{sucsec:analysis_LMext}
To extrapolate the available spectrograms towards giving an interpretation, in the next, we omit the softmax layer from the CNN and fed the images to the analyzer. The obtained output was normalized and the resulting images were stored using the \emph{flat} LRP rule which provided easily-interpretable results. Finally, the extracted images were superimposed over the original spectrograms. Moreover, relevance maps were averaged at the class level so as to obtain a global view of every class as illustrated in Fig \ref{fig:avg-spect}.

It should be mentioned that the implementation of layer-wise relevance propagation was based on the iNNvestigate library \cite{alber2018innvestigate} written in Python. The contribution of every single spectrogram in prediction interpretation is different for each class. In particular, it can be considered to be high for classes as siren and car horn, but poor for classes as gun shot where the output image is strongly influenced by the timing of the short sounds present in the 4 seconds samples; in such cases, it is preferable to focus on single spectrograms and not the averages.

\subsection{Interpretation results}

Looking at the average relevance maps, the primary observation is the direct relationship between assigned relevance and frequency content w.r.t each audio class. More specifically, by observing the averaged Mel-spectrogram maps, it is straightforward to relate a set of narrow frequency bands with specific classes. We argue that such relationships correspond well with human perception of the sounds. Taking as examples siren and car horn sound events, which are non-natural sounds, we can see how the red high relevance area is quite narrow due to specifications regarding sound production of the emitters. Moreover, dog bark sounds are typically restricted in a specific frequency band, which might be related to the actual pitch of barks.

For more complex sounds it is difficult to give a direct interpretation based solely on the relevance maps. Such examples include air conditioner, jackhammer, engine and drilling images, where relevance is distributed in wider parts of the spectrum. This might indicate that various frequency ranges have to be analyzed in order to accurately analyze such urban sounds.

Despite this, there are several cases where we can notice that particular sections of the spectrogram are consistently emphasized more than others. One example is on the engine idling analysis: the lower frequency part is mostly considered by the model. The relevance present in the higher frequencies is significantly lower showing that during the classification of such sound events, low frequency ranges dominate model analysis. In general, this corresponds well with human sound understanding, where low frequency content, typically characterizing low RPM engines, allows us to distinguish it from other sounds.

Last but not least, the averaged relevance in the gunshot sounds case is not particularly informative due to the timing issue described in section \ref{sucsec:analysis_LMext}. However, when examining single samples, it is interesting to note the importance that the model assigned to the peak of the sound: gunshots are in fact characterized by a sharp, high amplitude sound, distributed over a wide frequency range. A representative example is illustrated in Fig.\ref{fig:gun_mel}.

Overall, we argue that such a relevance-based analysis could be very useful in understanding and interpreting model decisions. At the same time, they could provide insights regarding the feature extraction stage as they might indicate spectrum parts which are not useful for classification, and thus, could be discarded.

\section{Conclusion}
\label{sec:conclusions}
Wider acceptance of machine learning based systems in real-world applications requests transparent and reliable interpretation of their predictions via a  rigorous etiology framework. This work addressed the critical application of assisting audio impaired drivers by means of an urban sound classification model. In this direction, two CNNs were constructed using two diverse spectrogram representations. Towards demystifying the operation of such deep neural networks, we employed the layer-wise relevance propagation technique and highlighted several direct relationships between the relevance assigned by the networks to specific frequency content with their predictions.

In the future, we wish to 
\begin{inparaenum}[a)]
    \item analyze performance when irrelevant feature regions are not considered,
    \item integrate the model in a mobile application including the interpretation functionality, and
    \item adapt the present analysis to interpret the operation of heterogeneous deep nets addressing different applications.
\end{inparaenum}

\section*{Acknowledgment}
We gratefully acknowledge the support of NVIDIA Corporation
with the donation of the Titan V GPU used for this research. This work was carried out within the project entitled Advanced methods for sound and music computing funded by the Piano Sostegno alla Ricerca of University of Milan.


\bibliographystyle{IEEEbib}
\bibliography{BiblioReconstruction}
\balance

\end{document}